\documentclass[preprint]{aastex}

\received{}
\accepted{}
\journalid{}{}
\articleid{}{}

\def\insertplot#1#2#3#4#5#6#7{
\vskip 10pt\nobreak\hbox to \hsize{\hss\dimen0=#3in\hbox to #6\dimen0{%
\dimen0=#2in\vbox to #6\dimen0{\vss
\special{ps: plotfile #1}
\special{ps::[end]
  PGPLOT restore
}
}\hss}\hss}\vskip 10pt}

\newcommand{\ts}{\thinspace}
\newcommand{\simless}{\mathbin{\lower 3pt\hbox
     {$\rlap{\raise 5pt\hbox{$\char'074$}}\mathchar"7218$}}}
\newcommand{\simgreat}{\mathbin{\lower 3pt\hbox
     {$\rlap{\raise 5pt\hbox{$\char'076$}}\mathchar"7218$}}}

\newcommand{\about}    {$\sim$\ts}
\newcommand{\aboutless}{$\simless$\ts}

\newcommand{\msun}{\ts M$_\odot$}

\newcommand{\etal}{et~al.}
\newcommand{\mdot}{$M{\raise 1.5ex\hbox{\hskip-6pt$\mathchar"201$}\kern0.2em}_{acc}$ }

\newcommand{\cmg}{\ts cm$^{2}$~g$^{-1}$}
\newcommand{\mjybeam}{\ts mJy\ts beam$^{-1}$}
\newcommand{\mjy}{\ts mJy}

\begin{document}

\title{Constraints on the Circumstellar Disk Masses in the IC~348 Cluster}

\author{John M. Carpenter}
\affil{California Institute of Technology, 
       Department of Astronomy, MS 105-24, \\ Pasadena, CA 91125; 
       email: jmc@astro.caltech.edu}

\begin{abstract}

A $5.2'\times5.2'$ region toward the young cluster IC~348 has been imaged
in the millimeter continuum at $4.0''\times4.9''$ resolution with the OVRO 
interferometer to a RMS noise level of 0.75\mjybeam\ at 98~GHz. The data are 
used to constrain the circumstellar disk masses in a cluster environment at an 
age of \about 2~Myr.  The mosaic encompasses 95 known members of the IC~348 
cluster with a stellar mass distribution that peaks at 0.2-0.5\msun. None of 
the stars are detected in the millimeter continuum at an intensity level of 
3$\sigma$ or greater.  The mean observed flux for the ensemble of 95 stars is 
$0.22\pm0.08$\mjy. Assuming a dust temperature of 20~K, a mass opacity 
coefficient of $\kappa_o = $ 0.02\cmg\ at 1300\micron, and a power law index 
of $\beta=1$ for the particle emissivity, these observations imply that the 
3$\sigma$ upper limit to the disk mass around any individual star is 
0.025\msun, and that the average disk mass is $0.002\pm0.001$\msun. The 
absence of disks with masses in excess of 0.025\msun\ in IC~348 is different 
at the \about 3$\sigma$ confidence level from Taurus, where \about 14\% of the 
stars in an optically selected sample have such disk masses. Compared with the 
minimum mass needed to form the planets in our solar system (\about 0.01\msun),
the lack of massive disks and the low mean disk mass in IC~348 suggest either
that planets more massive than a few Jupiter masses will form infrequently 
around 0.2-0.5\msun\ stars in IC~348, or that the process to form such 
planets has significantly depleted the disk of small dust grains on time 
scales less than the cluster age of \about 2~Myr.

\end{abstract}


\section{Introduction}

High resolution imaging at optical, near-infrared, and millimeter-wave 
wavelengths have conclusively demonstrated that circumstellar disks are 
ubiquitous around young solar-mass stars \citep{Odell96,Padgett99,Dutrey96}. 
The long-held notion that planets 
may frequently form in these disks has been dramatically confirmed with the 
increasing number of planets detected around nearby main-sequence stars 
\citep{Marcy00}. While a consensus has not yet been reached on how planets 
form in circumstellar disks, current theories propose that gas-giant planets 
form either as gravitational instabilities in an accretion disk 
\citep{Boss01}, or from the more gradual build up of planetesimals to form a 
rocky core followed by accretion of gaseous material \citep{Pollack96}. In 
both class of models, the accretion disk lifetime sets an important time scale 
on the planet formation process by constraining the time available for a 
gravitational instability to develop or the rocky core to form before the gas 
and dust are dissipated.

Circumstellar accretion disks around young stars are commonly inferred based 
on the presence of infrared emission in excess of the stellar photosphere. In 
many stars the excess infrared emission from 1-100\micron\ can be modeled 
with a geometrically thin, optically thick accretion disk \citep{Lynden74},
although inner disk holes, flaring, and other refinements to this basic model 
are often needed to reproduce the observations in detail 
\citep{Adams88,Calvet91,Chiang97}. Detailed spectral energy distributions from 
the near- to far-infrared have been compiled for only a few nearby star 
forming regions (e.g. Strom \etal~1989; Wilking, Lada, \& Young~1989), but  
$J$-, $H$-, $K$-, and $L$-band imaging studies provide an efficient means to 
search for near-infrared excesses in hundreds of stars (often in clusters) of 
various ages. Recent results indicate that at least 50\% of solar mass stars 
at an age of \about 1~Myr exhibit a near-infrared excess, but that the 
percentage decreases to \aboutless 10\% for ages 3-10~Myr 
\citep{Strom89,Haisch01b}. These results are consistent with the 
notion that the lifetime of accretion disks around most solar mass stars is 
\about 3-10~Myr.

Strictly speaking, a near-infrared excess is diagnostic of only hot dust with
temperatures of \about 1000~K. For a solar mass star, these temperatures are
found only within \about 0.1~AU of the star. Most of the dust mass, and
presumably the majority of planet formation, will be located at larger radii 
and have temperatures too cool to radiate in the near-infrared.  Further, some 
disks are thought to have inner holes such that stars without near-infrared 
excesses may still contain substantial disk masses. (Although 10\micron\ 
observations suggest that the transition from an optically thick to an 
optically thin disk occurs on time scales of \about 0.3~Myr, in which case 
near-infrared observations may in fact provide an indirect probe of the larger
scale disk; Skrutskie \etal~1990, see also Haisch, Lada, \& Lada 2001a). An 
additional limitation of 
near-infrared studies is that the emission is optically thick and provides 
little meaningful constraints on the disk mass (see, e.g., Wood \etal~2002).
Longer wavelength observations are needed to establish the disk lifetime at 
large orbital radii and to provide measures on how disk masses evolve with 
time.

Millimeter and submillimeter continuum emission are the best available tracers
of dust in the cool, outer disk. The most comprehensive continuum surveys 
to date have been toward young stars in Taurus \citep{Beckwith90,OB95,Motte01}, 
$\rho$~Oph \citep{Andre94,N98,Motte98}, Lupus \citep{N97}, Chamaeleon~I 
\citep{Henning93}, and Serpens \citep{Testi98}. These observations have shown 
that at an age of \about 1~Myr, \about 20-30\% of the stars possess a 
circumstellar disk with a mass greater than the minimum mass of the solar 
nebula (\about 0.01\msun; Weidenschilling 1977, Hayashi 1981), and that the 
median disk mass is 
\aboutless 0.004\msun. Further insights into disk evolution can be obtained by
using rich clusters as a tool to empirically measure the mass of disks as a
function of stellar mass and age in much the same manner near-infrared studies 
have used clusters to establish the frequency of inner accretion disks 
\citep{Haisch01b}. Continuum surveys of clusters are also important 
since an appreciable fraction of stars in molecular clouds are found in rich 
clusters \citep{Lada91,Carp00}, and the high ultraviolet radiation fields 
produced by O-stars in clusters may influence disk evolution 
\citep{Johnstone98,Storzer99,Scally01}. 

The main observational challenge to studying clusters in the millimeter 
continuum is that high angular resolution is required to resolve the stars and 
distinguish disk emission from the more extended molecular cloud. The only 
rich cluster surveyed to date by interferometric techniques has been the Orion 
Nebula Cluster \citep{Mundy95,Bally98}. However, these observations were 10-50 
times less sensitive to dust mass than the observations of Taurus and 
$\rho$~Oph and did not place stringent limits on the amount of circumstellar 
mass around individual stars.  Given the limitations in inferring disk 
evolution from near-infrared excesses, it is important to conduct millimeter 
continuum observations of additional star forming regions in order to 
understand the evolution of disk masses at young stellar ages when planets are 
in their formative stages.

To further measure the circumstellar disk masses in different star 
forming environments, I obtained $\lambda$3mm continuum observations of the 
young cluster IC~348 with the OVRO millimeter wave interferometer. Based on 
statistical analysis of near-infrared star counts, the IC~348 cluster 
contains \about 400 stars distributed over a \about $20'\times20'$ region
($1.9~\rm{pc}\times1.9~\rm{pc}$ at the assumed distance of 320~pc;
de Zeeuw \etal~1999; Herbig~1998), with about half the stars located within the 
central $7'\times7'$ region \citep{Lada95,Carp00}. To date, \about 200 stars 
and brown dwarfs have been individually identified as likely cluster members 
based on their spectroscopic and photometric properties 
\citep{Herbig98,Luhman98,Luhman99,Najita00}, and 115-173 likely members have
been detected with x-rays \citep{PZ01}. A cluster age of \about 2~Myr 
(Luhman 1999; see also Herbig~1998)
has been inferred by placing the low mass stars on the HR diagram and using 
theoretical pre-main-sequence evolutionary tracks from \citet{DM98}. 
Near-infrared imaging surveys of the cluster show that \about 65\% of the 
stars contain a $K-L$ excess characteristic of an optically thick accretion 
disk \citep{Lada95,Haisch01a}. By mosaicking the central $5.2'\times5.2'$ 
region ($0.48~\rm{pc}\times0.48~\rm{pc}$) of the IC~348 cluster in the 
millimeter continuum, I investigate the circumstellar disk masses in 95 known 
cluster members.

The OVRO observations of the IC~348 cluster are described in 
Section~\ref{obs}. The stellar properties of the cluster members within the 
OVRO mosaic boundaries are reviewed in Section~\ref{cluster}, and constraints 
on the circumstellar disk masses are derived in Section~\ref{constraints}. 
Section~\ref{discussion} compares these results with disk masses in Taurus and 
the Orion Nebula Cluster, and discuses the implications for planet formation. 
The conclusions are summarized in Section~\ref{summary}.

\section{OVRO Observations}
\label{obs}

A $5.2'\times5.2'$ region toward the IC~348 cluster was mosaicked in the
$\lambda$3mm continuum using the OVRO millimeter-wave interferometer
between September 2001 and January~2002. Continuum data were recorded
simultaneously in four, 1~GHz wide continuum channels centered at 96.48~GHz, 
97.98~GHz, 100.98~GHz, and 102.48~GHz. The phase and amplitude calibrator
was J0336+323 ($\alpha$,$\delta$ = 03:36:30.10,+32:18:29.3 J2000), which is
located 1.7\arcdeg\ from the cluster center. The data were calibrated using 
the OVRO MMA data reduction package \citep{Scoville93}. The adopted continuum 
flux for J0336+323, calibrated by observing Neptune and/or Uranus, was 
1.60~Jy.  The RMS dispersion in the measured flux for J0336+323 is 0.10~Jy as
computed from observations on 12 separate nights in which both the calibrator 
and the planets were observed.

\begin{figure}
\plotone{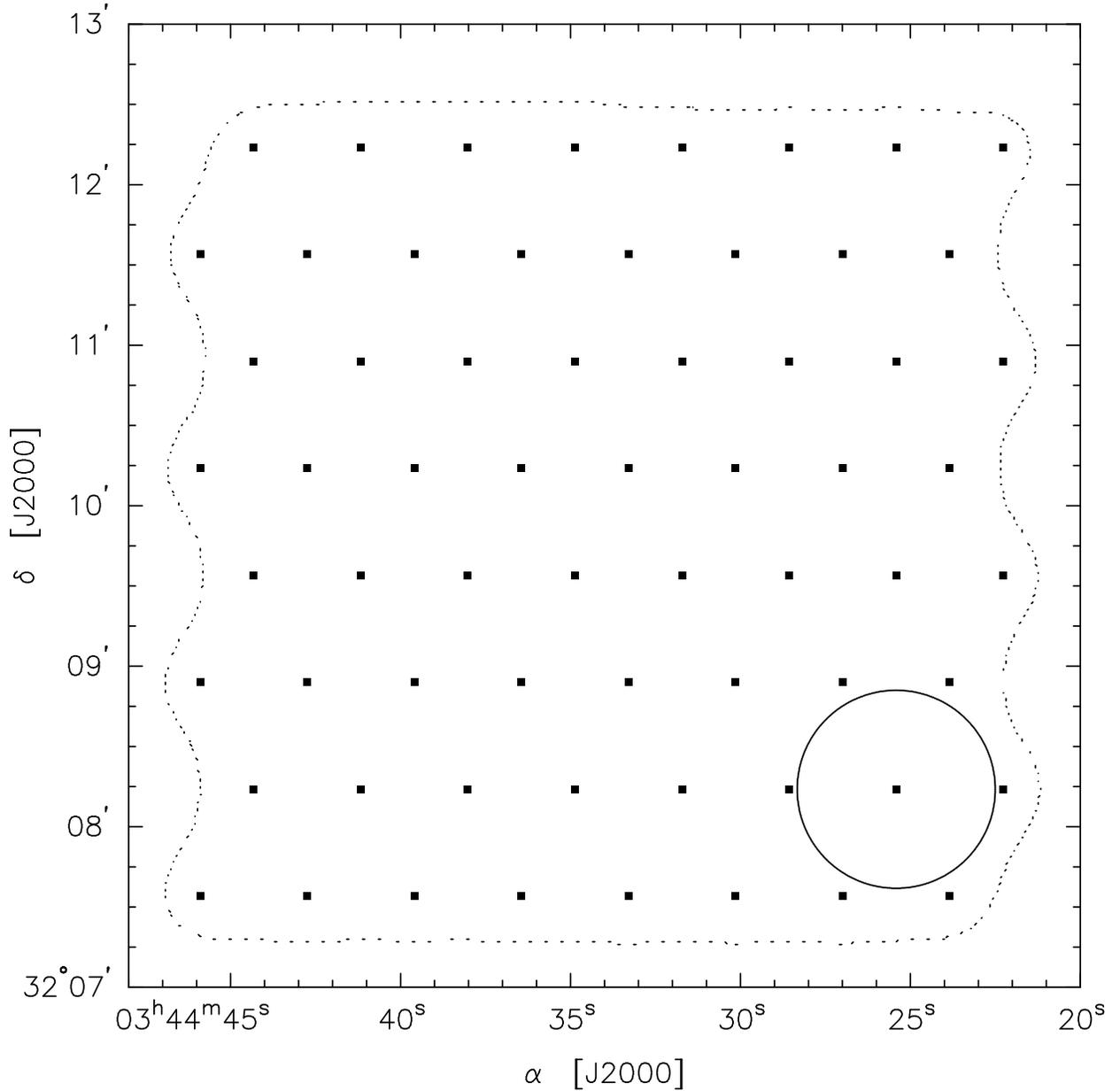}
\caption{
   Schematic of the IC~348 mosaic made with OVRO. The solid squares mark the 
   64 pointing centers used to create the mosaic. The pointing centers are 
   separated by 40$''$ along a row, with adjacent rows also separated by 
   40$''$. For comparison, the circle in the lower right corner
   shows the FWHM beam size of a single OVRO antenna (72\arcsec) at the 
   observed frequency 98~GHz. The dotted curve shows the extent of the OVRO
   mosaic (\about $5.2'\times5.2'$) at the unit gain boundary. 
   The synthesized beam size is $4.0''\times4.9''$.
\label{fig:gain}
}
\end{figure}

The mosaic consists of 64 pointing centers arranged in a $8\times8$ grid as 
shown in Figure~\ref{fig:gain}. The pointing centers along a row are separated 
by 40$''$ (compared to the primary FWHM beam size of 72$''$). Adjacent rows 
are also separated by 40$''$ but with pointing centers shifted by 20$''$ in
right ascension. The mosaicking sequence consisted of obtaining a 3 minute 
integration on the phase calibrator followed by observing 5 positions in the 
mosaic for 3 minutes each. This sequence was repeated until all 64 positions 
in the mosaic were observed. The mosaic was observed in the OVRO compact, low, 
equatorial, and high configurations both to increase the UV sampling and to 
decrease the RMS noise. The top few rows in the mosaic were often observed at 
high airmass with lower sensitivity. Additional observations of these rows 
were obtained to provide more uniform sensitivity across the mosaic.

The 64 individual points were formed into an image using the mosaicking 
routines in MIRIAD. The data were averaged using natural weighting to
maximize sensitivity, resulting in a beam size of $4.0''\times 4.9''$. 
Unit gain was achieved over a \about $5.2'\times5.2'$ area as shown by the 
dotted curve in Figure~\ref{fig:gain}. The average RMS noise within the unit 
gain region is 0.75~\mjybeam\ and is spatially uniform over the mosaic to 
within 10\%. 

\section{The IC 348 Cluster}
\label{cluster}

Before using the OVRO observations to constrain the circumstellar disk masses 
in the IC~348 cluster, I review the stellar properties of the IC~348 cluster 
members located within the mosaic boundaries. The stellar and substellar 
membership list of the IC~348 cluster was compiled from the spectroscopic 
observations by \citet{Herbig98}, \citet{Luhman98}, and \citet{Luhman99}, and 
the narrow band imaging survey by \citet{Najita00}. To ensure a reliable list
of cluster members, only sources brighter than $K_s$=14.5 were considered 
cluster members for this study, as the field star contamination becomes 
appreciable at fainter magnitudes \citep{Luhman98,Najita00}. In total, 95 
cluster members brighter than $K_s$=14.5 have been identified within the
OVRO mosaic. Astrometry for these sources was adopted from 2MASS. The 
astrometric uncertainty (\aboutless 0.2$''$) is 20 times less than the 
synthesized beam size.

Stellar masses and ages for 92 of the 95 cluster members have been estimated 
using the \citet{DM98} theoretical pre-main-sequence evolutionary tracks and 
published spectroscopic and photometric data as described in \citet{Lynne02}. 
Three sources do not have mass estimates since optical photometry is not 
available, or in one case, the spectral type could not be determined since the 
source is possibly a Class~I object \citep{Luhman98}. Figures~\ref{fig:mass} 
and \ref{fig:age} show histograms of the inferred stellar masses and ages 
respectively. (Also shown in these figures are the stellar mass and age 
histograms for a subsample of the Taurus population as discussed in 
Section~\ref{discussion}.) The stellar masses range from 0.03\msun\ to 
4.2\msun\ with a broad peak in the distribution at about 0.2-0.5\msun. The 
mean age of the stellar population is \about 2~Myr. The properties of this 
subsample agree well with the properties of the entire IC~348 population 
identified so far \citep{Herbig98,Luhman98,Luhman99}.

\begin{figure}
\epsscale{1.00}
\insertplot{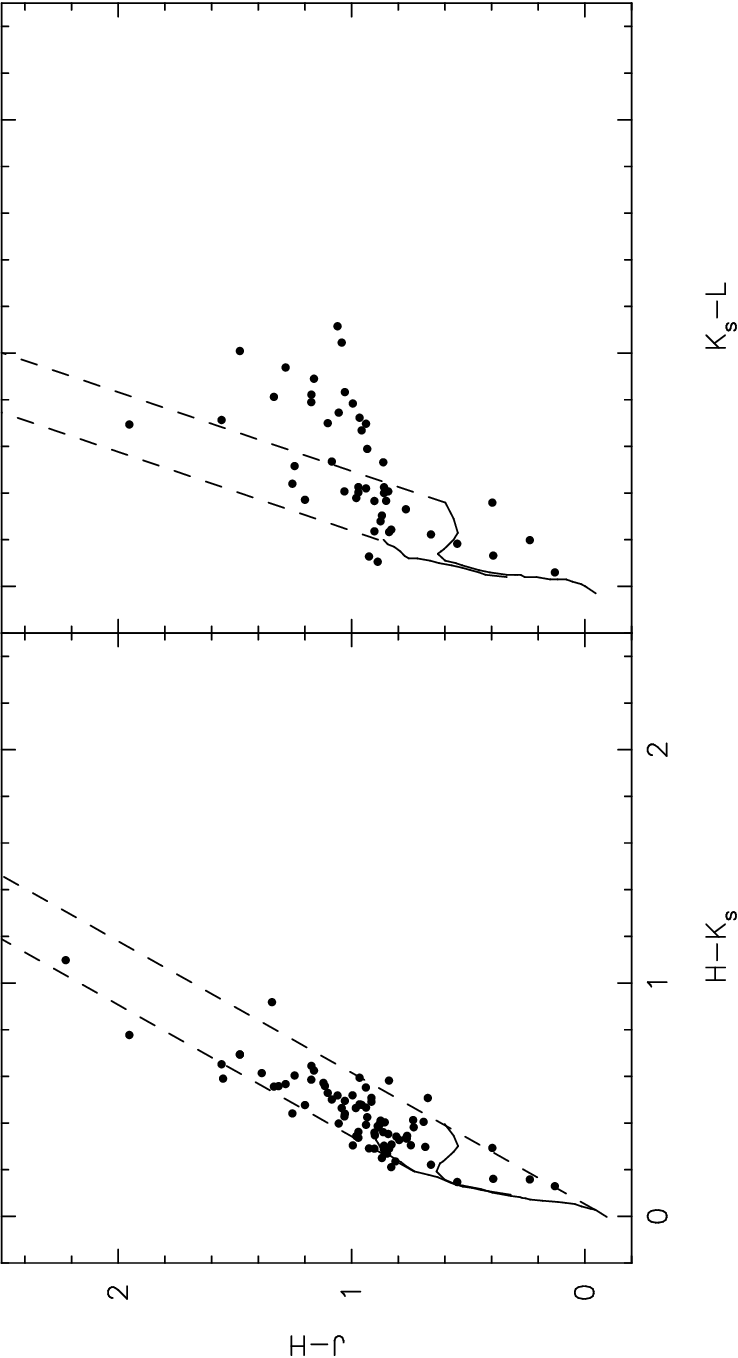}{6.8}{8.1}{-1}{0.8}{0.8}{1}
\caption{
  {\it Left}: $J-H$ vs. $H-K_s$ color-color diagram for 82 IC~348 members in 
  the OVRO mosaic that have photometry in the 2MASS database without any error 
  flags. 
  {\it Right}: $J-H$ vs. $K_s-L$ color-color diagram for 47 IC~348 members in 
  the OVRO mosaic with 2MASS photometry and $L$-band photometry from 
  \citet{Haisch01a}. In each panel, the solid curves represent the locus of 
  main-sequence and giant stars \citet{BB88} and the dashed line is the 
  interstellar reddening vector \citep{Cohen81}, where the $J-H$ and $H-K$ 
  colors have been transformed into the 2MASS photometric system 
  \citep{Carp01}. Four stars exhibit an apparent $H-K_s$ excess, but in each
  instance, the magnitude of the excess is less than the 1$\sigma$ photometric
  uncertainties and can be attributed to photometric noise. In the $J-H$ vs. 
  $K_s-L$ diagram, 18 stars show an apparent $K_s-L$ excess, of which 14 
  have an excess larger than the estimated photometric uncertainties.
  Therefore, a minimum of 15\% of the 95 IC~348 cluster members within the 
  OVRO mosaic contain a detectable $K_s-L$ excess indicative of an optically 
  thick circumstellar disk.
  \label{fig:ccd}
}
\end{figure}

To investigate the frequency distribution of inner accretion disks in this 
subsample of the IC~348 cluster, Figure~\ref{fig:ccd} shows the $J-H$ vs. 
$H-K_s$ and $J-H$ vs. $K_s-L$ color-color diagrams. The $J$, $H$, and $K_s$ 
photometry are from 2MASS, and the $L$-band photometry is from 
\citet{Haisch01a}. Only sources with no processing flags in the 2MASS database 
are shown. In Figure~\ref{fig:ccd}, the solid curves represent the locus of 
main-sequence and giant stars \citep{BB88} and the dashed line is the 
interstellar reddening vector \citep{Cohen81}, where the $J-H$ and $H-K$ 
colors have been transformed into the 2MASS photometric system \citep{Carp01}.
Of the 82 sources in the $J-H$ vs. $H-K_s$ diagram, 4 have near-infrared 
colors to the right of the reddening vectors that are consistent 
with an infrared excess. However, for each of these four stars, the magnitude
of the apparent $H-K_s$ excess is less than the 1$\sigma$ photometric 
uncertainties and therefore may be attributed to photometric noise. The 
fraction of stars with an infrared excess is higher in the $J-H$ vs. $K_s-L$ 
diagram since $L$-band shows larger contrast between the disk and photospheric 
emission. Of the 47 stars with available $J$, $H$, $K_s$, and $L$-band 
photometry, 18 (38\%) contain an apparent $K_s-L$ excess. \citet{Haisch01a} do 
not report $L$-band photometric uncertainties for individual sources. 
Based on the distribution of stars blueward of the un-reddened main-sequence
in the $J-H$ vs. $K_s-L$ diagram (a region which cannot be populated except 
from photometric noise and photometric variability between the time of the 
2MASS and $L$-band observations), I estimate that \about 4 sources may 
have an apparent $K_s-L$ excess due to noise. Therefore, a minimum of \about 
15\% of the 95 IC~348 members within the OVRO mosaic have near-infrared colors 
indicative of an optically thick, inner accretion disk. By comparison,
\citet{Haisch01a} estimate that the \about 65\% of the stars in the IC~348 
cluster brighter than $K$=12 have a $K-L$ excess.

\section{Constraints on Disk Masses}
\label{constraints}

\begin{figure}
\epsscale{0.90}
\insertplot{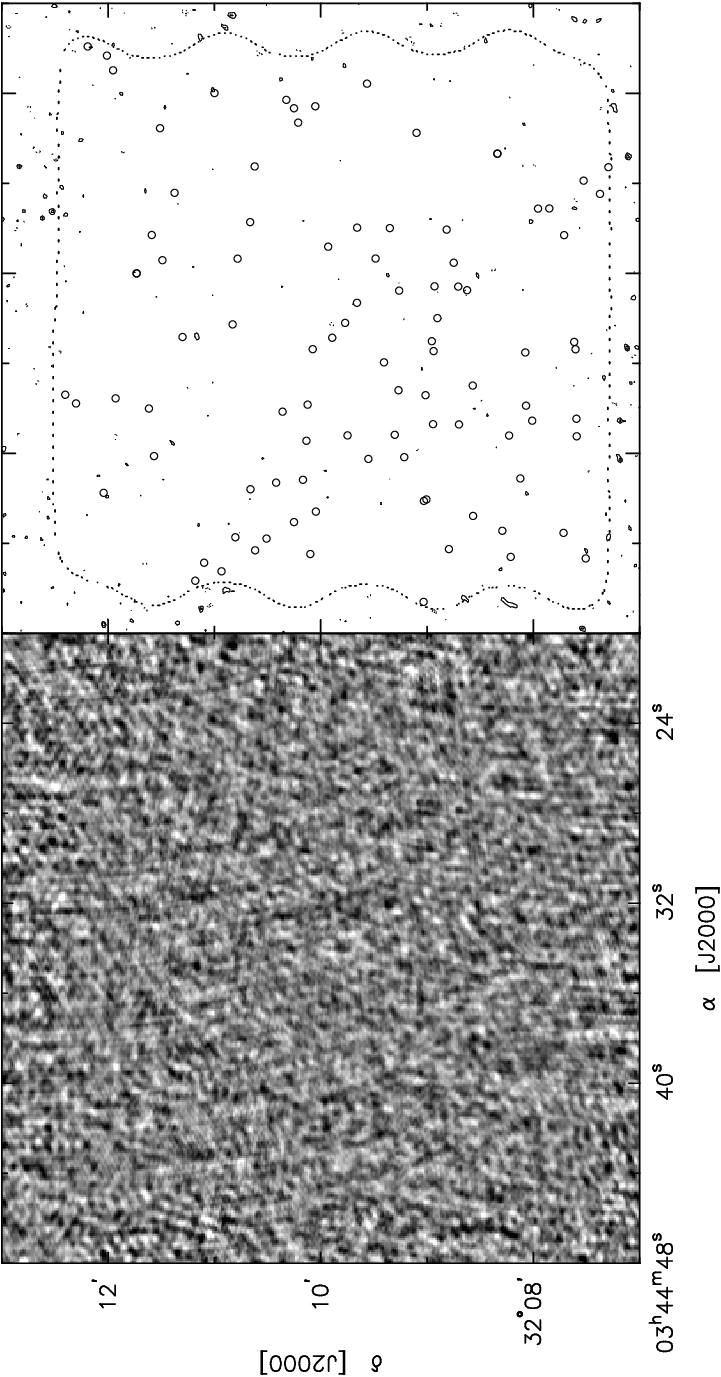}{6.8}{8.1}{-1}{0.8}{0.8}{1}
\caption{
  {\it Left}: Grayscale image of the $\lambda$3mm continuum emission toward
  IC~348. Darker regions represent bright intensities.
  {\it Right}: Contour map of the OVRO mosaic. Contours begin at 3$\sigma$ 
  above the RMS noise of 0.75\mjybeam\ with increments of 1$\sigma$.
  The dotted boundary that encompasses the image shows the unit gain 
  boundary of the mosaic (see Fig.~\ref{fig:gain}). 
  Open circles represent 95 probable members of the IC~348 cluster within the 
  field of view of the OVRO mosaic that have been identified from spectroscopy 
  \citep{Herbig98,Luhman98,Luhman99} and narrow band imaging \citep{Najita00}. 
  This figure shows that none of the known IC~348 stars were detected in the 
  $\lambda$3mm continuum at the 3$\sigma$ noise level or greater.
  \label{fig:map}
}
\end{figure}

\begin{figure}
\plotone{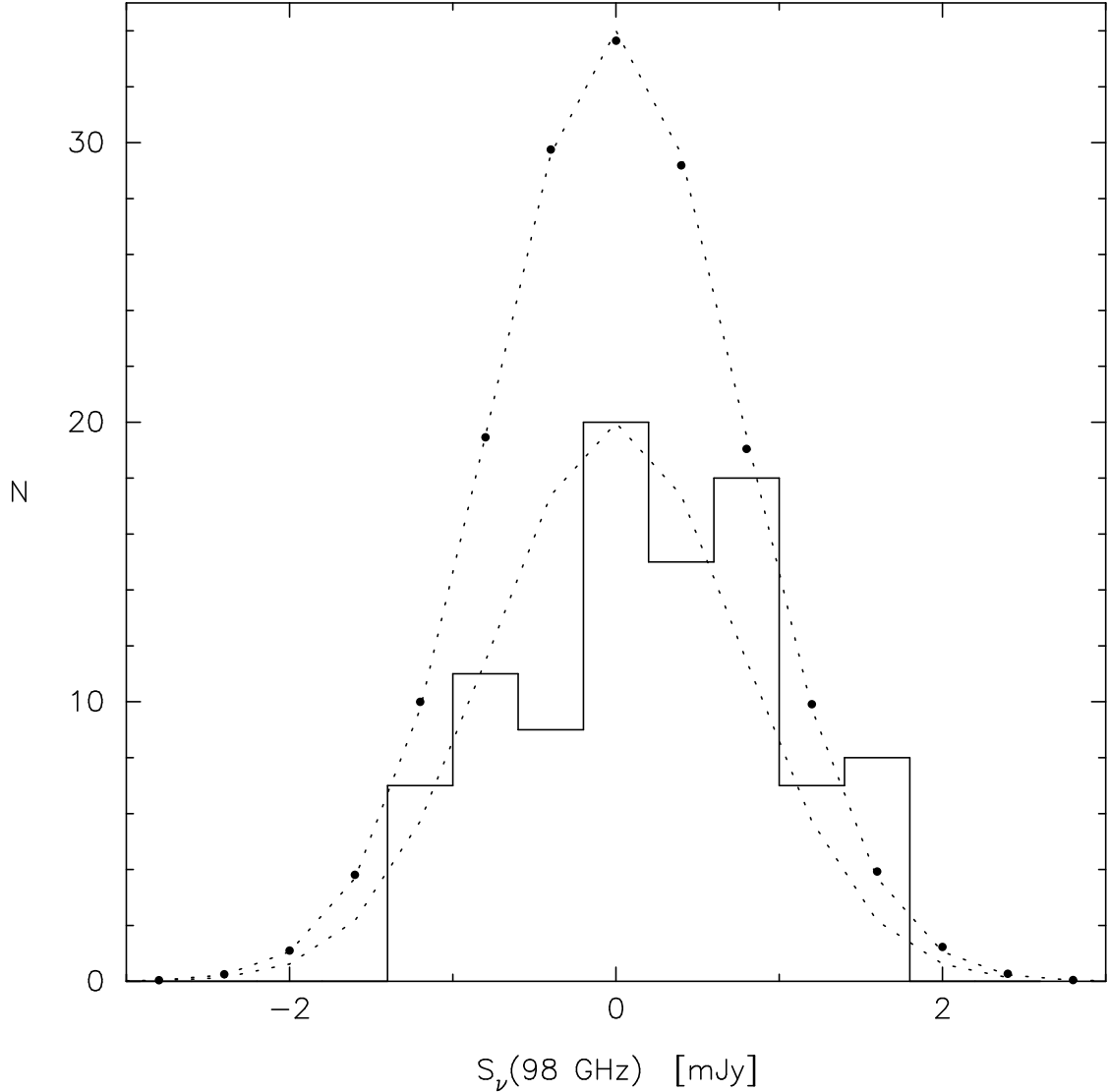}
\caption{
  Frequency distribution of the observed flux densities at 98~GHz over the 
  entire OVRO mosaic (solid circles) and toward the 95 IC~348 members
  (histogram). The frequency distribution indicated by the solid circles have
  been arbitrarily scaled by 1/600. The dotted curves through the circles and 
  through the histogram show the expected distribution of fluxes in the two 
  samples for gaussian noise with a dispersion of 0.75\mjybeam, which is the
  average noise in the OVRO mosaic. This figure shows that the fluxes over the 
  entire mosaic are consistent with gaussian noise. The mean flux observed 
  toward the 95 IC~348 members is $0.22\pm0.08$\mjy, where the uncertainty is 
  the standard deviation of the mean. 
  \label{fig:flux}
}
\end{figure}

Figure~\ref{fig:map} shows gray scale and contour maps of the OVRO mosaic. In 
the contour map, the dotted curve shows the unit gain boundary of the mosaic, 
and the open circles indicate the positions of the 95 known members of the 
IC~348 cluster within the unit gain boundary. The contours begin at 3$\sigma$ 
above the RMS noise of 0.75\mjybeam\ with increments of 1$\sigma$. Visual 
inspection of the contour map shows that none of the known IC~348 cluster 
members have been detected in the $\lambda$3mm continuum at the 3$\sigma$ 
level or greater. Over the entire mosaic, the brightest observed flux is only 
3.9$\sigma$ above the noise, and the frequency distribution of fluxes are 
consistent with gaussian noise as shown by the solid circles in 
Figure~\ref{fig:flux}. Therefore there are no definitive continuum detections 
in the mosaic.

Circumstellar disks in Taurus typically have diameters of \aboutless 300~AU
as measured in the $\lambda$3mm continuum \citep{Dutrey96}.
Assuming that the disks in IC~348 are similar, the disks can be considered
point sources at the resolution of these observations (1600~AU$\times$1300~AU) 
when computing upper limits to the continuum flux. A histogram of the observed 
fluxes toward the 95 cluster members is shown in Figure~\ref{fig:flux}. The 
dotted line through the histogram shows the expected flux distribution 
for gaussian noise given the observed RMS noise of 0.75\mjybeam. The observed 
mean flux toward the cluster members is $0.22 \pm 0.08$\mjy, where the 
uncertainty represents the standard deviation of the mean of the 95 cluster 
members. These results indicate that while the flux distribution is consistent 
with gaussian noise, there is a positive bias to the average flux at the 
2.8$\sigma$ confidence level.

The OVRO continuum observations can be used to place constraints on the 
circumstellar disk masses. In reality, most disks will contain a range of 
dust temperatures with decreasing temperatures with increasing radial distance
from the star. The data obtained here do not permit sophisticated modeling of 
the disk structure, however, and a single dust temperature was assumed.  
Following \citet{Hildebrand83}, the disk mass (i.e. sum of the dust and gas 
components) can then be estimated as 
$M_{disk}  = {S_\nu\:D^2\over \kappa_\nu\:B_\nu(T_{dust})}$,
where $\kappa_\nu = \kappa_o({\nu\over\nu_o})^\beta$ is the mass opacity
coefficient, $\beta$ parameterizes the frequency dependence of $\kappa_\nu$,
$S_\nu$ is the observed flux, $T_{dust}$ is the dust temperature, and 
$B_\nu(T_{dust})$ is the Planck function. For consistency with prior studies, 
I assumed $\beta$=1.0 and $\kappa_o$ = 0.02\cmg\ at 1300\micron\ 
\citep{Beckwith90}. The value of $\kappa_o$, and consequently the disk masses,
is uncertain by at least a factor of three \citep{Beckwith90}. The power law 
index $\beta$ may be as low as \about 0.5 on average in circumstellar disks 
\citep{Beckwith91,Mannings94}. If this lower value of $\beta$ is 
more appropriate for circumstellar disks, the assumptions adopted here will 
overestimate the disk masses by 50\%.

\begin{figure}
\plotone{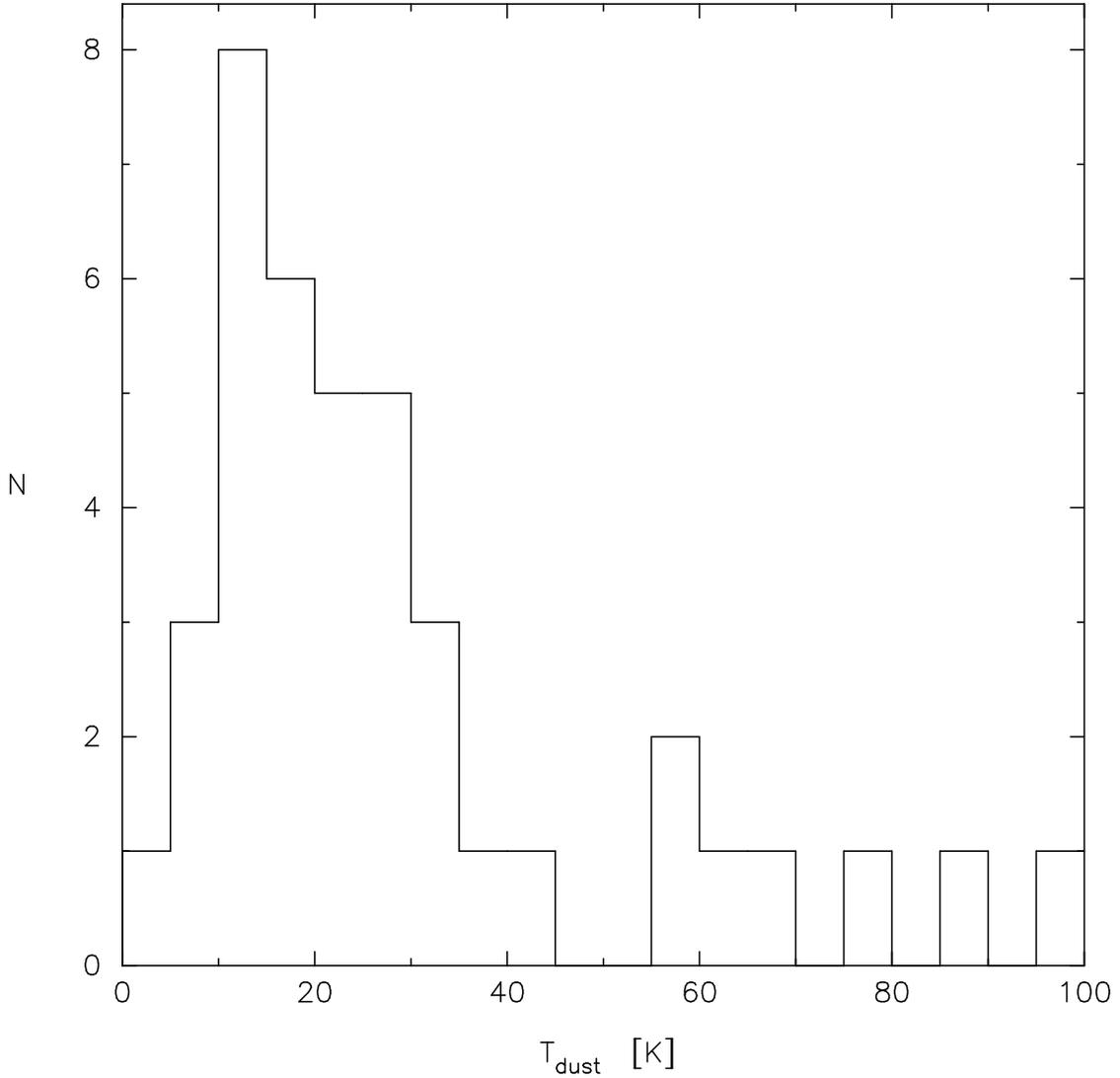}
\caption{
  Histogram of the effective dust temperatures of circumstellar disks in
  Taurus-Auriga. The effective dust temperature represents the dust 
  temperature that, when combined with the observed $\lambda$1.3mm fluxes,
  reproduces the disk masses derived by fitting power-law distributions for the
  dust temperature and mass surface density to the observed spectral energy 
  distribution of young stars in Taurus-Auriga \citep{Beckwith90,OB95}.
  Most stars have effective dust temperatures of \about 20~K, which was used
  to convert the observed OVRO $\lambda$3mm fluxes to disk masses. Four 
  stars have effective dust temperatures greater than 100~K and are not shown 
  in this figure.
  \label{fig:tdust}
}
\end{figure}

The appropriate dust temperature for calculating the disk masses was 
estimated using the circumstellar disk models from \citet{Beckwith90} and 
\citet{OB95}. These studies fitted radial power-law distributions for
the dust temperature and disk mass surface density to the observed infrared 
and millimeter spectral energy distribution for well-studied stars in 
Taurus-Auriga. From the disk mass derived by these model fits, I computed a 
single component dust temperature for each star that reproduces the disk mass 
given the observed $\lambda$1.3mm flux. A histogram of these `effective''
dust temperatures is shown in Figure~\ref{fig:tdust}. (Four stars with 
effective temperatures greater than 100~K are not shown in this figure.)
Of the 44 stars in Taurus-Auriga with model disk masses from 
\citet{Beckwith90} and \citet{OB95}, the median dust temperature derived in 
this manner is \about 20~K. If this dust temperature is then assumed for all 
of the Taurus stars, the model-derived dust masses can be reproduced to within 
a factor of 2 for 60\% of the stars, and within a factor of 3 for 70\% of the 
stars. (For sources with effective dust temperatures substantially greater
than 20~K, the single component dust model will overestimate the disk mass
since $M_{\rm dust} \propto T_{\rm dust}^{-1}$). Assuming that the disk 
temperatures in the IC~348 cluster are not vastly different than in Taurus, 
adopting a dust temperature of 20~K, while simplistic, should reproduce the 
disk masses within a factor of a few for the majority of the stars.
For these assumptions, the 3$\sigma$ upper limit to the circumstellar
disk mass around an individual star in the IC~348 cluster is 0.025\msun.
Treating the 95 cluster members as an ensemble, the observed average flux 
corresponds to an average disk mass of $0.002\pm0.001$\msun.

\section{Discussion}
\label{discussion}

The limits on the disk masses in the IC~348 cluster can be compared to disk 
masses in other star forming regions as a step toward determining the 
evolution of disk masses as a function of time and environment (e.g. 
``clustered'' vs.  ``isolated'' star formation). The Taurus star forming 
region provides a meaningful comparison since it has been 
thoroughly studied in the millimeter continuum (Beckwith \etal~1990; Osterloh
\& Beckwith~1995; see also Skinner, Brown, \& Walter~1991 and Henning 
\etal~1998), and the stellar population has an age similar to IC~348 as shown
below. Since the IC~348 sample has been identified largely based on 
spectroscopy, only stars in Taurus with spectral types and sufficient 
photometric information to place the stars on the HR diagram were used for 
this comparison. From the database of spectral types and photometry compiled 
from the literature by \citet{Lynne02}, 117 stars were identified in Taurus 
that met these criteria and have a published $\lambda$1.3mm continuum flux 
measurement. The main source of incompleteness in the Taurus sample are deeply 
embedded IRAS sources which are not detected at optical wavelengths
\citep{KH95}, and recently identified low mass stars and brown dwarfs 
\citep{Briceno99,LR98} that have not yet been surveyed in the millimeter 
continuum.

\begin{figure}
\epsscale{0.55}
\plotone{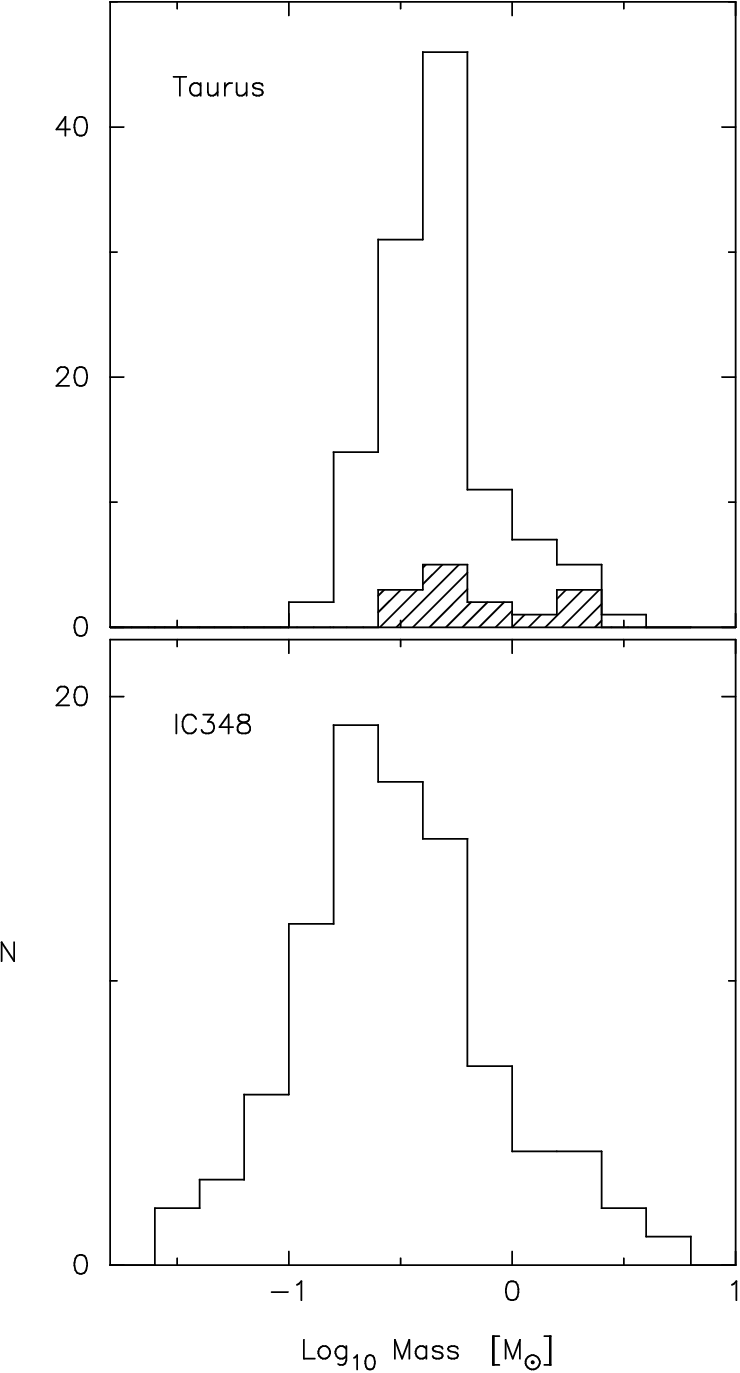}
\caption{
  {\it Bottom:} Histogram of the stellar masses for the IC~348 cluster members 
  located within the OVRO mosaic. Three probable IC~348 members within the 
  OVRO mosaic did not have sufficient data to infer their stellar properties 
  and are not shown in this figure.
  {\it Top:} Histograms of the stellar masses for a comparison sample of stars 
  in the Taurus molecular cloud that have available optical spectroscopy, 
  photometry, and submillimeter continuum observations. The open histogram
  represents all stars in the comparison sample, and the hatched histogram
  shows the stellar mass distribution for sources that have a disk mass 
  greater than the 3$\sigma$ detection limit of 0.025\msun\ for the IC~348 
  observations.  The stellar masses in both IC~348 and Taurus have been 
  inferred by placing the stars in an HR diagram using the database compiled 
  by \citet{Lynne02} and the \citet{DM98} pre-main-sequence evolutionary 
  tracks.
  \label{fig:mass}
}
\end{figure}

\begin{figure}
\epsscale{0.55}
\plotone{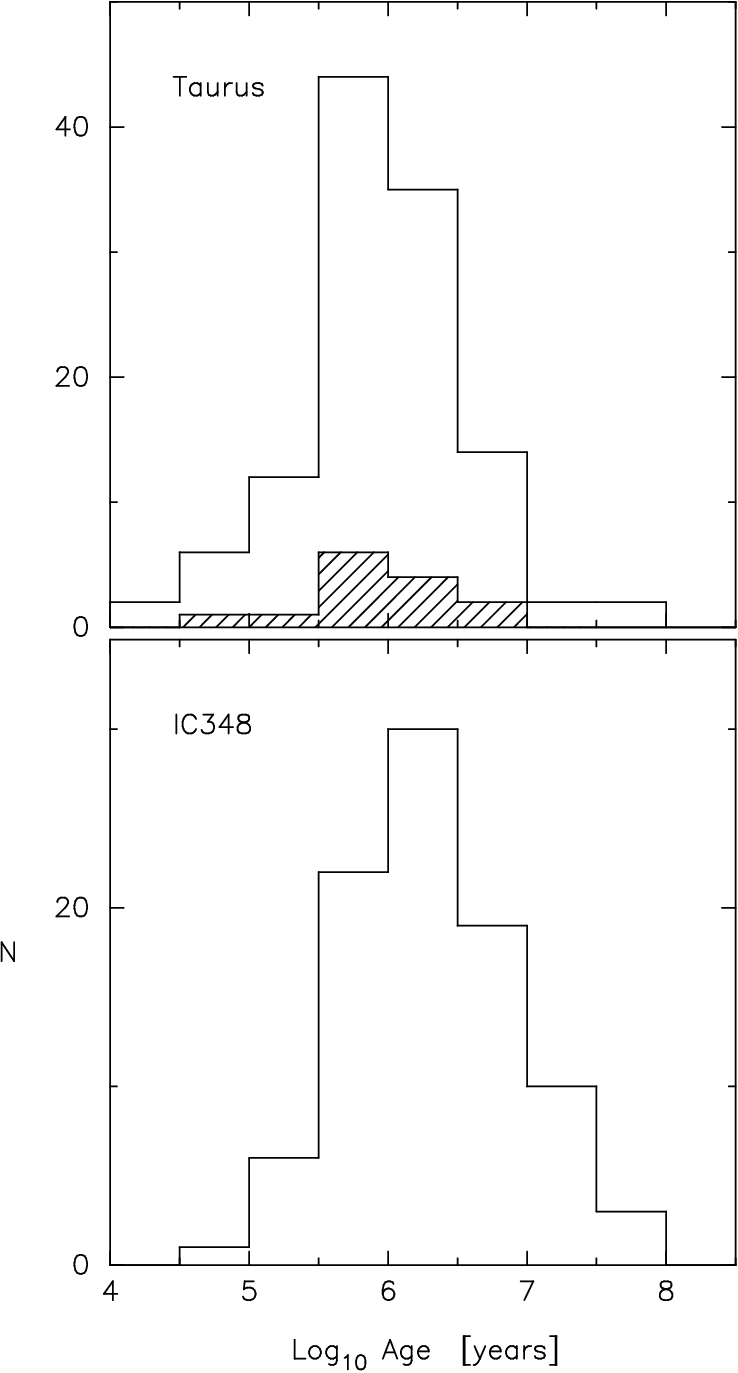}
\caption{
  {\it Bottom:} Histogram of the stellar ages for the IC~348 cluster members 
  within the OVRO mosaic.
  {\it Top:} Histograms of the stellar ages for a comparison sample of stars 
  in the Taurus molecular cloud as described in Fig.~\ref{fig:mass}.
  \label{fig:age}
}
\end{figure}

The open histograms in the top panels of Figures~\ref{fig:mass} and 
\ref{fig:age} show the mass and age distributions respectively for the Taurus 
comparison sample. (The bottom panels show the analogous distributions for
the IC~348 sample as already discussed.) The mean stellar mass of the Taurus 
sample is log$_{10}$ (mass/\msun) = $-0.33$ with a dispersion of 0.27. By 
comparison, the IC~348 sample has a mean log$_{10}$ stellar mass of $-0.53$ 
with a 
dispersion of 0.44. Therefore, the stellar mass distribution in IC~348
is skewed toward smaller stellar masses compared to the Taurus sample. 
Similarly, the mean stellar age of the Taurus sample is log$_{10}$ (age/yr) = 
6.0 with dispersion of 0.6, while the mean log$_{10}$ age and dispersion for 
IC~348 
are 6.3 and 0.6 respectively. Thus the stellar ages for IC~348 are skewed 
toward older values than stars in Taurus. However, the uncertainty in the 
distances to the star forming regions, which is at least \about 10\%,
\citep{Kenyon94,DZ99}, will produce a \about 0.09 dex uncertainty in the mean 
age \citep{Hartmann01}. At best then the difference in the mean ages between 
Taurus and IC~348 are significant at the \about 2$\sigma$ confidence level.

Disk masses for stars in Taurus were re-computed from published 
$\lambda$1.3mm fluxes and using the isothermal dust model adopted here for 
consistency. The hatched histograms in Figures~\ref{fig:mass} and \ref{fig:age} 
show the distribution of stellar masses and ages for stars in Taurus that 
have disk masses in excess of 0.025\msun, which is the 3$\sigma$ upper limit 
to the disk masses toward individual stars in IC~348. Figure~\ref{fig:mass} 
shows that in this Taurus sample, such disks masses are found only around 
stars more massive than 0.27\msun. While the lack of such massive disks around 
lower mass stars may simply be due to the relative lack of sources, only stars 
more massive than 0.27\msun\ are used in the following comparison between 
Taurus and IC~348. In the Taurus sample, 14 of the 99 stars with stellar 
masses greater than 0.27\msun\ have disk masses in excess of 0.025\msun. In 
IC~348, none of the 48 stars in the same stellar mass range have such disk 
masses. Using the two-tailed Fisher Exact Test, the probability that the 
frequency distribution of massive disks in 
IC~348 and Taurus have been drawn from the same parent population is 0.0028. 
This probability is independent of the assumed values of $\kappa_o$ and 
$T_{\rm dust}$ used to compute the disk masses to the extent that the disk 
properties in the two regions are similar. However, if the value of $\beta$ is 
as low as 0.5 on average \citep{Beckwith91,Mannings94} as opposed to the 
assumed value of $\beta$=1, then the disk masses in IC~348 will be 
systematically overestimated compared to Taurus, and the difference in the 
frequency of massive disks will be enhanced. Thus despite the similar stellar 
ages of the star forming regions, the IC~348 cluster lacks the massive disks 
found in the Taurus at the \about 3$\sigma$ confidence level. The cause of 
this apparent difference cannot be due to an photoevaporation of the disks in
IC~348 from a high ultraviolet radiation field (e.g. Johnstone, Hollenbach, 
\& Bally~1998) since the spectral type of the most massive cluster member in 
IC~348 is B5 \citep{Luhman98}. Whether or not the difference in disk masses 
can be attributed to the affects of a ``cluster'' vs. ``isolated'' environment 
remains speculative given the limited sample of star forming regions, although 
continuum surveys of additional clusters are possible with current 
interferometers.

The IC~348 results can also be compared to the disk masses in Orion~Nebula
Cluster using the 86~GHz continuum survey by \citet{Mundy95}. Again using a 
20~K dust temperature for consistency with the assumptions adopted here, the 
3$\sigma$ upper limit to individual disk masses in Orion is 0.17\msun. 
\citet{Mundy95} derived a 95\% upper limit to the characteristic flux at 
86~GHz of 1.1\mjy, which corresponds to a disk mass of 0.03\msun. Thus the 
results from \citet{Mundy95} exclude the presence of massive disks in Orion, 
but do not set as stringent limits as the IC~348 observations for disk masses 
in cluster environments. Similarly, \citet{Bally98} derived a 3$\sigma$ upper 
limit of 0.015\msun\ to the disk mass around 5 proplyds in the Orion Nebula 
cluster for an assumed dust temperature of 50~K. If a 20~K dust temperature is 
instead adopted, the corresponding upper limit is 0.047\msun.

Finally, the IC~348 results can be contrasted with the minimum disk mass 
needed to form the Solar System. Summing the mass contained in the planets 
and assuming an interstellar gas to dust ratio, the minimum mass of the
primitive solar nebula is \about 0.01\msun\ \citep{W77,Hayashi81}. This minimum 
mass reflects primarily the mass needed to form a Jupiter-mass planet. The 
actual mass of the primitive solar nebula may of course be higher depending on 
the efficiency in converting dust and gas in the primitive solar nebula into 
planets. The upper limit to the typical disk mass in IC~348 is comparable then
to the minimum mass of the solar nebula, and the average mass is \about 5 
times lower than minimum mass of the solar nebula. Given that the disk masses 
are uncertain by at least a factor of 3 due the mass opacity coefficient 
($\kappa_o$), these results suggest either that the formation of planets with 
masses greater than a few Jupiter masses is relatively rare around 
0.2-0.5\msun\ stars in IC~348, or that the formation of gas-giant planets 
has significantly depleted the disk of small dust grains on a time scale of 
\aboutless 2~Myr, the mean age of the IC~348 cluster.

\section{Summary}
\label{summary}

I present OVRO observations of a $5.2'\times5.2'$ region toward the IC~348 
stellar cluster that has been imaged in the 98~GHz continuum to a RMS noise 
level of 0.75\mjybeam\ at $4.0''\times4.9''$ resolution. A total of 
95 known members of the IC~348 cluster brighter than $K_s$=14.5 are located 
within the mosaicked 
region. The stellar masses range from 0.03\msun\ to 4.2\msun\ with a peak in 
the distribution between 0.2\msun\ and 0.5\msun. At least 15\% of these stars 
are surrounded by an optically thick, inner accretion disk as evidenced by a 
$K_s-L$ excess (see Haisch, Lada, \& Lada~2001a). The OVRO observations are 
used to place constraints on the circumstellar disk masses and provide a 
snapshot of disk evolution in a cluster environment at an age of \about 2~Myr 
\citep{Luhman99}.

None of the 95 IC~348 members within the OVRO mosaic were detected at the
3$\sigma$ level or greater. The mean flux toward the 95 stars is 
$0.22\pm0.08$\mjy, indicating the observed fluxes are consistent with random
noise although with a slight bias toward positive fluxes. The millimeter-wave 
continuum fluxes were converted into disk masses (the sum of the gas and dust
components) assuming a dust temperature of 20~K, a mass opacity coefficient of 
0.02\cmg\ at 1300\micron, and a power law index for the particle emissivity of 
$\beta$=1. For any individual source, the 3$\sigma$ upper limit to the 
circumstellar disk mass is 0.025\msun. For the 95 sources as an ensemble, the 
average disk mass is $0.002\pm0.001$\msun.

The constraints placed on the disk masses in the IC~348 cluster were compared
to the disk masses around stars in the Taurus molecular cloud. A subsample of 
the Taurus population was identified that has available optical spectroscopy 
and photometry as well as published millimeter continuum observations. The 
stellar mass range was restricted to stars with masses in excess of 0.27\msun\ 
since that is the lowest mass star in the Taurus sample that has an observed 
disk mass in excess of 0.025\msun. In the Taurus sample, 14 of the 99 stars 
with stellar masses $\ge$0.27\msun\ have disk masses in excess of 0.025\msun, 
while in IC~348, none of the 48 stars in the same stellar mass range have such 
disk masses. The probability that the frequency distribution of massive disks 
in IC~348 and Taurus have been drawn from the same parent population is 
0.0028, and are thus different at the \about 3$\sigma$ confidence level. These 
results suggest that the IC~348 stellar population lacks the massive disks 
that characterizes \about 14\% of the Taurus population despite the fact that 
the two regions have similar ages \citep{Lynne02}. Whether or not the 
differences between the frequency of massive disks between Taurus and IC~348 
is related to ``isolated'' versus ``clustered'' star forming environments 
awaits observations of additional star forming regions.

The implications of the IC~348 observations for planet formation can be 
placed in context with the minimum mass of the primitive solar nebula
(\about 0.01\msun; Weidenschilling 1977, Hayashi 1981) needed to form our 
Solar System. The IC~348 results suggest that if an optically thick accretion 
disk is the necessary seed to form gas-giant planets, then either few stars in 
IC~348 with stellar masses between 0.2\msun\ and 0.5\msun\ will form planets 
more massive than a few Jupiter masses, or the process to form such massive
planets has significantly depleted the disk of small dust grains by the age 
of the IC~348 cluster, or \about 2~Myr.

\acknowledgements

JMC acknowledges support from Long Term Space Astrophysics Grant NAG5-8217 and 
the Owens Valley Radio Observatory, which is supported by the National Science 
Foundation through grant AST-9981546.
This publication makes use of data products from the Two Micron
All Sky Survey, which is a joint project of the University of Massachusetts
and the Infrared Processing and Analysis Center, funded by the National
Aeronautics and Space Administration and the National Science Foundation.
2MASS science data and information services were provided by the InfraRed
Science Archive (IRSA) at IPAC.

\clearpage

\clearpage

\clearpage

\clearpage

\clearpage

\clearpage

\clearpage

\clearpage

\end{document}